\documentclass[a4paper,12pt,twoside]{cesj}
\usepackage{amssymb}
 \newcommand{\sech}{\rm \ sech \,}
 \newcommand{\cosech}{\rm \ cosech \,}
\newcommand{\cosec}{\rm \ cosec \,}
 \newcommand{\cotan}{\rm \ cotan \,}
\newcommand{\cotanh}{\rm \ cotanh \,}
\year{}
\issue{}
\rightmark{Nicolae Cotfas}
\leftmark{Nicolae Cotfas}

\title{Shape invariant hypergeometric type 
       operators with application to quantum mechanics}

\author{Nicolae Cotfas \email{E-mail: ncotfas@yahoo.com}}

\institute{Faculty of Physics, University of Bucharest, PO Box 76 - 54, 
Bucharest, Romania}

\received{}
\revised{}

\abstract{ A hypergeometric type equation satisfying certain conditions
defines either a finite or an infinite system of orthogonal polynomials.     
The associated special functions are eigenfunctions of some shape invariant
operators. These operators can be analysed together and the mathematical 
formalism we use can be extended in order to define other shape invariant operators.
All the considered shape invariant operators are directly related 
to Schrodinger type equations. }



\begin{document}
\firstpage{1}
\maketitle \setcounter{page}{1}%

\section{Introduction}

Many problems in quantum mechanics and
mathematical physics lead to equations of the type
\begin{equation}\label{hypeq}
\sigma (s)y''(s)+\tau (s)y'(s)+\lambda y(s)=0 \label{eq}
\end{equation}
where $\sigma (s)$ and $\tau (s)$ are polynomials of at most second
and first degree, respectively, and $\lambda $ is a constant. 
These equations are usually called {\em equations of hypergeometric
type} \cite{NUS}, and each of them can be reduced to the self-adjoint form 
\begin{equation}
[\sigma (s)\varrho (s)y'(s)]'+\lambda \varrho (s)y(s)=0 
\end{equation}
by choosing a function $\varrho $ such that 
$(\sigma \varrho )'=\tau \varrho $.

The equation (\ref{hypeq}) is usually considered on an interval $(a,b)$,
chosen such that 
\begin{equation}\begin{array}{r}
\sigma (s)>0\qquad {\rm for\ all}\quad s\in (a,b)\\
\varrho (s)>0\qquad {\rm for\ all}\quad s\in (a,b)\\
\lim_{s\rightarrow a}\sigma (s)\varrho (s)
=\lim_{s\rightarrow b}\sigma (s)\varrho (s)=0.
\end{array}
\end{equation}
Since the form of the equation (\ref{eq}) is invariant under a 
change of variable $s\mapsto cs+d$, it is sufficient to analyse the cases
presented in table 1.
Some restrictions are to be imposed to $\alpha $, $\beta $ in
order the interval $(a,b)$ to exist. 
The equation (\ref{hypeq}) defines an infinite sequence of orthogonal polynomials
in the case $\sigma (s)\in \{ 1,\ s,\ 1-s^2\}$, and a finite one 
in the case $\sigma (s)\in \{ s^2-1,\ s^2,\ s^2+1\}$.\\

\begin{table}[htbp]
\begin{center}
\begin{tabular}{c|cllc}
\hline
$\sigma (s)$ & $\tau (s)$  & $\varrho (s)$ & $\alpha ,\beta $ &  
$(a,b) $\\
\hline \hline
$1$ & $\alpha s+\beta $ & ${\rm e}^{\alpha s^2/2+\beta s}$ & $\alpha <0$
& $\mathbb{R}$\\
$s$ & $\alpha s+\beta $ & $s^{\beta -1} {\rm e}^{\alpha s}$ & 
$\alpha <0$, $\beta >0$& $(0,\infty )$\\ 
$1-s^2$  & $\alpha s+\beta $ & $(1+s)^{-(\alpha -\beta )/2-1}
(1-s)^{-(\alpha +\beta )/2-1}$ & 
$\alpha <\beta <-\alpha $ & $(-1,1)$\\
$s^2-1$  & $\alpha s+\beta $ & $(s+1)^{(\alpha -\beta )/2-1}
(s-1)^{(\alpha +\beta )/2-1}$ &
$-\beta <\alpha <0$ & $(1,\infty )$\\
$s^2$  & $\alpha s+\beta $ & $s^{\alpha -2}{\rm e}^{-\beta /s}$ & $\alpha <0$, $\beta >0$ &
$(0,\infty )$\\
$s^2+1$  & $\alpha s+\beta $ & $(1+s^2)^{\alpha /2-1}{\rm e}^{\beta \arctan s}$ & 
$\alpha <0$ & $\mathbb{R}$\\
\hline
\end{tabular}
\end{center}
\caption{The main cases}
\end{table}

The literature discussing special function theory and its application to mathematical
and theoretical physics is vast, and there are a multitude of different conventions
concerning the definition of functions. A unified approach 
is not possible without a unified definition for the associated special functions.
In this paper we define them as 
\begin{equation}
 \Phi _{l,m}(s)=\left(\sqrt{\sigma (s)}\right)^m\, \frac{{\rm d}^m}{{\rm d}s^m}\Phi _l(s)
\end{equation}
where  $\Phi _l$ are the orthogonal polynomials defined by equation  (\ref{hypeq}). 
The table 1 allows 
one to pass in each case from our parameters $\alpha $, $\beta $ to the parameters
used in different approach.

In \cite{C,C1} we presented a systematic study of the Schr\"odinger
equations exactly solvable in terms of associated special functions. 
In the present paper, based on the factorization method \cite{CKS,IH} and certain results of 
Jafarizadeh and Fakhri \cite{JF}, we extend our unified formalism  by adding other shape invariant operators.

\section{Orthogonal polynomials}

Let $\tau (s)=\alpha s+\beta $ be a fixed polynomial, and let
\begin{equation}
\lambda _l\!=-\frac{\sigma ''(s)}{2}l(l-1)-\tau '(s)l
\!=-\frac{\sigma ''}{2}l(l-1)-\alpha \,l
\end{equation}
for any $l\in \mathbb{N}$. It is well-known \cite{NUS} that for $\lambda =\lambda _l$,
the equation (\ref{hypeq}) admits a polynomial solution 
$\Phi _l=\Phi _l^{(\alpha ,\beta )}$ of at most $l$ degree
\begin{equation} \label{eq3}
\sigma (s) \Phi _l ''+\tau (s) \Phi _l '+\lambda _l\Phi _l=0.
\end{equation}
If the degree of the polynomial $\Phi _l$ is $l$ then it satisfies the
Rodrigues formula \cite{NUS}
\begin{equation}
\Phi _l(s)=\frac{B_l}{\varrho (s)}\frac{{\rm d}^l}{{\rm d}s^l}[\sigma ^l(s)\varrho (s)]
\end{equation}
where $B_l$ is a constant. Based on the relation 
\begin{equation} 
\begin{array}{l} 
\{ \ \delta \in \mathbb{R}\ |\ 
\lim_{s\rightarrow a}\sigma (s)\varrho (s)s^\delta  
=\lim_{s\rightarrow b}\sigma (s)\varrho (s)s^\delta =0 \ \}\\[2mm]
\mbox{}\qquad \qquad =\left\{
\begin{array}{lll}
[0,\infty ) & {\rm if} & \sigma (s)\in \{ 1,\ s,\ 1-s^2\}\\[2mm] 
[0,-\alpha ) & {\rm if} & \sigma (s)\in \{ s^2-1,\ s^2,\ s^2+1\}
\end{array} \right.  
\end{array} 
\end{equation}
one can prove \cite{C1,NUS} that the system of polynomials $\{\Phi _l\ |\ l<\Lambda \}$, where
\begin{equation}
\Lambda \!=\!\left\{ \begin{array}{lcl}
\infty & {\rm for} & \sigma (s)\in \{ 1,\ s,\ 1-s^2\}\\[2mm]
\frac{1-\alpha }{2} & { \rm for } & 
\sigma (s)\in \{ s^2\!-\!1,\ s^2,\ s^2\!+\!1\} 
\end{array}\right.
\end{equation}
is orthogonal with weight function $\varrho (s)$ in $(a,b)$. This means that
equation (\ref{hypeq}) defines an infinite sequence of orthogonal polynomials
\[ \Phi _0,\ \ \Phi _1,\ \ \Phi _2,\ ... \]
in the case $\sigma (s)\in \{ 1,\ s,\ 1-s^2\}$, and a finite one 
\[ \Phi _0,\ \ \Phi _1,\ \ ...,\ \ \Phi _L \]
with \ $L=\max \{ l\in \mathbb{N}\ |\ l<(1-\alpha )/2\}$
in the case $\sigma (s)\in \{ s^2-1,\ s^2,\ s^2+1\}$.

The polynomials $\Phi _l^{(\alpha ,\beta )}$ can be expressed (up to a multiplicative constant) in terms of the 
classical orthogonal polynomials as
\begin{equation}\label{classical}
  \Phi _l^{(\alpha ,\beta )}(s)=\left\{ \begin{array}{lcl}
{\bf H}_l\left(\sqrt{\frac{-\alpha }{2}}\, s-\frac{\beta }{\sqrt{-2\alpha }}\right)  
& {\mbox{}\quad {\rm in\ the\ case}\quad \mbox{}} & \sigma (s)=1\\[2mm]
{\bf L}_l^{\beta -1}(-\alpha s)  & {\rm in\ the\ case} & \sigma (s)=s\\[2mm]
{\bf P}_l^{(-(\alpha +\beta )/2-1,\ (-\alpha +\beta )/2-1)}(s)  & {\rm in\ the\ case} & \sigma (s)=1-s^2\\[2mm]
{\bf P}_l^{((\alpha -\beta )/2-1,\ (\alpha +\beta )/2-1)}(-s)  & {\rm in\ the\ case } & \sigma (s)=s^2-1\\[2mm]
\left(\frac{s}{\beta }\right)^l{\bf L}_l^{1-\alpha -2l}\left(\frac{\beta }{s}\right) 
& {\rm in\ the\ case} & \sigma (s)=s^2\\[2mm]
{\rm i}^l{\bf P}_l^{((\alpha +{\rm i}\beta )/2-1,\ (\alpha -{\rm i}\beta )/2-1)}({\rm i}s) 
& {\rm in\ the\ case} & \sigma (s)=s^2+1
\end{array} \right.
\end{equation}
where ${\bf H}_l$, ${\bf L}_l^p $ and ${\bf P}_l^{(p,q)}$ are the Hermite,
Laguerre and Jacobi polynomials, respectively. The relation (\ref{classical}) does not have a very simple form. 
In certain cases we have to consider the classical polynomials 
outside the interval where they are orthogonal or for complex values of parameters.

\section{Associated special functions. Shape invariant operators}

Let $l\in \mathbb{N}$, $l<\Lambda $, and let $m\in \{ 0,1,...,l\}$.
The functions
\begin{equation}\label{def}
\Phi _{l,m}(s)=\kappa ^m(s)\frac{{\rm d}^m}{{\rm d}s^m}\Phi _l(s) \qquad {\rm where}\qquad 
\kappa (s)=\sqrt{\sigma (s)}
\end{equation}  
are called the {\em associated special functions}. 
If we  differentiate (\ref{eq3}) $m$ times and then multiply 
the obtained relation by $\kappa ^m(s)$ then we get the equation
\begin{equation}\label{Hm}
{H}_m \Phi _{l,m}=\lambda _l\Phi _{l,m}
\end{equation}
where ${H}_m$ is the differential operator
\begin{equation} \label{defHm}
\begin{array}{l}
{H}_m =-\sigma (s) \frac{d^2}{ds^2}-\tau (s) \frac{d}{ds}
+\frac{m(m-2)}{4}\frac{(\sigma '(s))^2}{\sigma (s)}\\[2mm]   
\mbox{}\qquad \ \ 
 + \frac{m\tau (s)}{2}\frac{\sigma '(s)}{\sigma (s)}
-\frac{1}{2}m(m-2)\sigma ''(s)-m\tau '(s) .
\end{array}
\end{equation}

For each $m<\Lambda $, the special functions $\Phi _{l,m}$ with $m\leq l<\Lambda $ 
are orthogonal with respect to the scalar product 
\begin{equation} \label{scalarprod}
 \langle f,g\rangle 
=\int_a^b\overline{f(s)}\, g(s)\varrho(s)ds
\end{equation}
and the functions corresponding to consecutive values of $m$ are related 
through the raising/lowering operators \cite{C,C1}
\begin{equation}\begin{array}{l}
A_m=\kappa (s)\frac{d}{ds}-m\kappa '(s)\\[3mm]
A_m^+=-\kappa (s)\frac{d}{ds}-\frac{\tau (s)}{\kappa (s)}-(m-1)\kappa '(s)
\end{array}
\end{equation}
namely, 
\begin{equation}\label{AmAm+}
\begin{array}{l}
A_m\Phi _{l,m}=\left\{ \begin{array}{lll}
0 & {\rm for} & l=m\\
\Phi _{l,m+1} & {\rm for} & m<l<\Lambda 
\end{array} \right. \\[5mm]
A_m^+\Phi _{l,m+1}\!=\!(\lambda _l\!-\!\lambda _m)\Phi _{l,m}\ \  
{\rm for}\ \ 0\leq m<l< \Lambda .
\end{array}
\end{equation}
Up to a multiplicative constant 
\begin{equation}\label{philm}
\Phi _{l,m}(s)=\left\{ \begin{array}{lll}
\kappa ^l(s) & {\rm for} & m=l\\
\frac{A_m^+ }{\lambda _l-\lambda _m}
\frac{A_{m+1}^+ }{\lambda _l-\lambda _{m+1}}...
\frac{A_{l-1}^+ }{\lambda _l-\lambda _{l-1}}\kappa ^l(s) \quad & {\rm for} \quad & m<l
\end{array} \right.
\end{equation}
and the operators $H_m$ are shape invariant \cite{C,C1}
\begin{equation}\label{fact}
\begin{array}{ll}
{H}_m-\lambda _m=A_m^+A_m & A_m{H}_m={H}_{m+1}A_m \\
{H}_{m+1}-\lambda _m =A_mA_m^+\qquad  & {H}_mA_m^+=A_m^+{H}_{m+1}. 
\end{array}
\end{equation} 

The functions 
\begin{equation}
 \phi _{l,m}=\Phi _{l,m}/||\Phi _{l,m}||
\end{equation}
where $||f||=\sqrt{\langle f,f\rangle }$
are the {\em normalized associated special functions}. 
Since \cite{C,C1}
\begin{equation} \label{norm}
||\Phi _{l,m+1}||
=\sqrt{\lambda _l-\lambda _m}\, ||\Phi _{l,m}||\qquad  
\end{equation}
they satisfy the relations
\begin{equation}\label{relations} \begin{array}{l}
A_m\ \phi _{l,m}=\left\{ \begin{array}{lll}
0 & {\rm for} & l=m\\
\sqrt{\lambda _l-\lambda _m}\ \phi _{l,m+1} & {\rm for} & m<l<\Lambda 
\end{array} \right.\\[3mm]
A_m^+\ \phi _{l,m+1}=\sqrt{\lambda _l-\lambda _m}\ \phi _{l,m}
\quad {\rm for}\  0\leq m<l<\Lambda \\[3mm]
\phi _{l,m}=
\frac{A_m^+ }{\sqrt{\lambda _l-\lambda _m}}
\frac{A_{m+1}^+ }{\sqrt{\lambda _l-\lambda _{m+1}}}...
\frac{A_{l-1}^+ }{\sqrt{\lambda _l-\lambda _{l-1}}}\phi _{l,l}. 
\end{array}
\end{equation}

\section{Application to Schr\" odinger type equations}

It is well-known \cite{IH} that the equations ${H}_m \Phi _{l,m}=\lambda _l\Phi _{l,m}$
are directly related to certain Schr\' odinger type equations. 
If in equation satisfied by $\Phi _{l,m}$
\begin{equation} 
\begin{array}{l}
-\sigma (s) \frac{d^2}{ds^2}\Phi _{l,m}(s)-\tau (s) \frac{d}{ds}\Phi _{l,m}(s)
+\lbrack \frac{m(m-2)}{4}\frac{(\sigma '(s))^2}{\sigma (s)} \\[2mm]
\qquad + \frac{m\tau (s)}{2}\frac{\sigma '(s)}{\sigma (s)}  
-\frac{1}{2}m(m-2)\sigma ''(s)-m\tau '(s)\rbrack \Phi _{l,m}(s) =\lambda _l\Phi _{l,m}(s)
\end{array}
\end{equation}
we pass to a new variable $x=x(s)$ and a new function $\Psi _{l,m}(x)$ such that 
\begin{equation}\label{xofs}
\frac{dx}{ds}=\xi (s)\qquad \qquad \Phi _{l,m}(s)=\eta (s)\ \Psi _{l,m}(x(s))
\end{equation}
$\xi (s)\not=0$ and $\eta (s)\not=0$ for any $s\in (a,b)$, then we get the equation
\begin{equation} \label{sch}
\begin{array}{l}
-\sigma (s)\, \xi ^2(s)\, \ddot {\Psi }_{l,m}(x(s))-\lbrack \sigma (s)\xi '(s)+
2\sigma (s)\, \xi (s)\, \frac{\eta '(s)}{\eta (s)}\\[2mm]
\qquad \qquad +\tau (s)\xi (s)\rbrack \dot {\Psi }_{l,m}(x(s))
+V_m(s)\Psi _{l,m}(x(s)) =\lambda _l\, \Psi _{l,m}(x(s))
\end{array}
\end{equation}
where
\begin{equation} \begin{array}{l} 
V_m(s)=\frac{m(m-2)}{4}\frac{(\sigma '(s))^2}{\sigma (s)}   
 + \frac{m\tau (s)}{2}\frac{\sigma '(s)}{\sigma (s)}-\frac{1}{2}m(m-2)\sigma ''(s)\\[2mm] 
\quad \qquad \qquad \qquad \qquad -m\tau '(s)\!-\!\sigma (s)\frac{\eta ''(s)}{\eta (s)}
-\tau (s)\frac{\eta '(s)}{\eta (s)}
\end{array}
\end{equation}
and the dot sign means derivative with respect to $x$. For $\xi (s)$ and $\eta (s)$ satisfying the conditions
\begin{equation} \label{eta} 
\sigma (s)\, \xi ^2(s)=1\qquad \quad 
\sigma (s)\xi '(s)+2\sigma (s)\, \xi (s)\, \frac{\eta '(s)}{\eta (s)}+\tau (s)\xi (s)=0.
\end{equation}
which lead to 
\begin{equation}
\xi (s)=\pm \frac{1}{\kappa (s)}\qquad \qquad \eta (s)=\frac{1}{\sqrt{\kappa (s)\, \varrho (s)}}
\end{equation}
(up to a multiplicative constant), the equation (\ref{sch}) becomes
\begin{equation} \label{sch1}
-\ddot {\Psi }_{l,m}(x(s))+V_m(s)\Psi _{l,m}(x(s)) =\lambda _l\, \Psi _{l,m}(x(s)).
\end{equation}
Denoting by $s(x)$ the inverse of the function $(a,b)\longrightarrow (a',b'):s\mapsto x(s)$ we get  
\begin{equation}
\frac{ds}{dx}=\pm \kappa (s(x))\qquad \qquad 
\Psi _{l,m}(x)=\sqrt{\kappa (s(x))\, \varrho (s(x))}\, \Phi _{l,m}(s(x)).
\end{equation}
The equation (\ref{sch1}) is satisfied for any $s\in (a,b)$ if and only if 
\begin{equation}
-\ddot {\Psi }_{l,m}(x)+V_m(s(x))\Psi _{l,m}(x) =\lambda _l\, \Psi _{l,m}(x)
\qquad {\rm for\ any}\ \  x\in (a',b') 
\end{equation}
that is, if and only if $\Psi _{l,m}(x)$ is an eigenfunction of the
Schr\" odinger type operator 
\begin{equation} \label{calH}
\mathcal{H}_m=-\frac{d^2}{dx^2}+\mathcal{V}_m(x)\qquad {\rm where}\qquad \mathcal{V}_m(x)=V(s(x)).
\end{equation}

For each $m<\Lambda $ the functions $\Psi _{l,m}(x)$ with $m\leq l<\Lambda $ are orthogonal
\[ \begin{array}{l}\int_{a'}^{b'}\overline{\Psi }_{l,m}(x)\Psi _{k,m}(x)dx
=\int_a^b\overline{\Phi }_{l,m}(s(x))\Phi _{k,m}(s(x))\varrho (s(x))\, \left|\frac{ds}{dx}\right|\, dx\\[2mm] 
\qquad \qquad \qquad \qquad \quad =\int_a^b\overline{\Phi }_{l,m}(s)\Phi _{k,m}(s)\varrho (s)ds=0 
\end{array} \]
for $k\not= l$, and satisfy the relations
\begin{equation} \begin{array}{l}
  {\mathcal A}_m\Psi _{l,m}(x)=\left\{ \begin{array}{lll}
0 & {\rm for} & l=m\\
\Psi _{l,m+1} & {\rm for} & m<l<\Lambda 
\end{array} \right. \\[5mm]
{\mathcal A}_m^+ \Psi _{l,m+1}(x)=(\lambda _l-\lambda _m)\Psi _{l,m}(x)
\end{array}
\end{equation}
where 
\begin{equation}\label{tildeA+}
 \begin{array}{l}
{\mathcal A}_m=[\kappa (s)\varrho (s)]^{1/2}A_m[\kappa (s)\varrho (s)]^{-1/2}|_{s=s(x)}\\[2mm]
{\mathcal A}_m^+ =[\kappa (s)\varrho (s)]^{1/2}A_m^+ [\kappa (s)\varrho (s)]^{-1/2}|_{s=s(x)}
\end{array}
\end{equation}
are the operators corresponding to $A_m$ and $A_m^+ $. Particulary, we have \cite{IH}
\begin{equation} \label{relcalH}
\begin{array}{ll}
\mathcal{H}_m-\lambda _m=\mathcal{A}_m^+\mathcal{A}_m & 
\mathcal{A}_m\mathcal{H}_m=\mathcal{H}_{m+1}\mathcal{A}_m \\[2mm]
\mathcal{H}_{m+1}-\lambda _m =\mathcal{A}_m\mathcal{A}_m^+\qquad  & 
\mathcal{H}_m\mathcal{A}_m^+=\mathcal{A}_m^+\mathcal{H}_{m+1}. 
\end{array}
\end{equation} 
and
\begin{equation}  
\Psi _{l,m}(x)=
\frac{{\mathcal A}_m^+ }{\lambda _l-\lambda _m}
\frac{{\mathcal A}_{m+1}^+ }{\lambda _l-\lambda _{m+1}}...
\frac{{\mathcal A}_{l-2}^+ }{\lambda _l-\lambda _{l-2}}
\frac{{\mathcal A}_{l-1}^+ }{\lambda _l-\lambda _{l-1}}
\Psi _{l,l}(x)
\end{equation}  
for each $m\in \{ 0,1,...,l-1\}$.\\[5mm]
\begin{theorem}
If the change of variable $s=s(x)$ is such that $ds/dx=\pm \kappa (s(x))$ then
\begin{equation}\label{calA}
{\mathcal A}_m=\pm \frac{d}{dx}+W_m(x)\qquad \qquad {\mathcal A}_m^+ =\mp \frac{d}{dx}+W_m(x)
\end{equation}
and
\begin{equation}
\mathcal{V}_m(x)=W_m^2(x)\mp \dot W_m(x)+\lambda _m =\frac{\ddot \Psi _{m,m}(x)}{\Psi _{m,m}(x)}+\lambda _m 
\end{equation}
where $W_m(x)$ is the superpotential \cite{JF}
\begin{equation}\label{Wm}
W_m(x)=-\frac{\tau (s(x))}{2\kappa (s(x))}
- \left(m-\frac{1}{2}\right )\frac{d\kappa }{ds}(s(x))=\mp \frac{\dot \Psi _{m,m}(x)}{\Psi _{m,m}(x)}.
\end{equation}
\end{theorem}
\begin{proof*}
From $(\sigma \varrho )'=\tau \varrho $ and $ds/dx=\pm \kappa (s(x))$ we get
\[ \frac{\varrho '}{\varrho }=\frac{\tau }{\kappa ^2}-2\frac{\kappa '}{\kappa }\qquad \quad
\frac{d}{ds}=\pm \frac{1}{\kappa (s(x))}\frac{d}{dx} \]
whence (\ref{calA}). Since $\mathcal{A}_m\Psi _{m,m}=0$, from (\ref{calH}), (\ref{relcalH}) 
and (\ref{calA}) we obtain
\[ \pm \dot \Psi _{m,m}+W_m(x)\Psi _{m,m}=0\qquad \quad 
-\ddot \Psi _{m,m}+(V_m(x)-\lambda _m)\Psi _{m,m}=0. \]
\end{proof*}
The functions 
\begin{equation}
\psi _{l,m}(x)=\sqrt{\kappa (s(x))\, \varrho (s(x))}\, \phi _{l,m}(s(x)).
\end{equation}
corresponding to $\phi _{l,m}$ are normalized 
\[  \int_{a'}^{b'}|\psi _{k,m}(x)|^2 dx
=\int_a^b|\phi _{k,m}(s(x))|^2\varrho (s(x))\, \left|\frac{ds}{dx}\right|\, dx 
=\int_a^b|\phi _{k,m}(s)|^2\varrho (s)ds=1 \]
and satisfy the relations
\begin{equation}\label{relations} \begin{array}{l}
\mathcal{A}_m\ \psi _{l,m}=\left\{ \begin{array}{lll}
0 & {\rm for} & l=m\\
\sqrt{\lambda _l-\lambda _m}\ \psi _{l,m+1} & {\rm for} & m<l<\Lambda 
\end{array} \right.\\[5mm]
\mathcal{A}_m^+\ \psi _{l,m+1}=\sqrt{\lambda _l-\lambda _m}\ \psi _{l,m}
\quad {\rm for}\  0\leq m<l<\Lambda \\[5mm]
\psi _{l,m}=
\frac{\mathcal{A}_m^+ }{\sqrt{\lambda _l-\lambda _m}}
\frac{\mathcal{A}_{m+1}^+ }{\sqrt{\lambda _l-\lambda _{m+1}}}...
\frac{\mathcal{A}_{l-1}^+ }{\sqrt{\lambda _l-\lambda _{l-1}}}\psi _{l,l}. 
\end{array}
\end{equation}

\noindent {\bf Particular cases} \cite{CKS,D,JF}.
Let $\alpha _m=-(2m+\alpha -1)/2$, \ 
$\alpha '_m=(2m-\alpha -1)/2.$ 
\begin{enumerate}

\item {\it Shifted oscillator}\\  
In the case $\sigma (s)=1$, the change of variable 
$\mathbb{R}\longrightarrow \mathbb{R}:\, x\mapsto s(x)=x$
leads to
\begin{equation}
\begin{array}{l}
W_m(x)=-\frac{\alpha x+\beta }{2}\\
\mathcal{V}_m(x)=\frac{(\alpha x+\beta )^2}{4}+\frac{ \alpha }{2}+\lambda _m
\end{array}
\end{equation}
where $\lambda _m=-\alpha m$.

\item {\it Three-dimensional oscillator}\\
In the case $\sigma (s)=s$, the change of variable
$(0,\infty )\longrightarrow (0,\infty ):\, x\mapsto s(x)=x^2/4$
leads to 
\begin{equation}
\begin{array}{l}
W_m(x)=-\frac{\alpha }{4}x-\left(\beta +m-\frac{1}{2}\right) \frac{1}{x}\\
\mathcal{V}_m(x)=\frac{\alpha ^2}{16}x^2
+\left(\beta +m-\frac{1}{2}\right)\left(\beta +m-\frac{3}{2}\right)\frac{1}{x^2}
+\frac{\alpha }{2}(\beta +m)+\lambda _m
\end{array}
\end{equation}
where $\lambda _m=-\alpha m$.

\item {\it P\"{o}schl-Teller type potential}\\  
In the case $\sigma (s)=1-s^2$, the change of variable
$(0,\pi )\longrightarrow (-1,1):\, x\mapsto s(x)=\cos x$
leads to 
\begin{equation}
\begin{array}{l}
W_m(x)=\alpha '_m {\cotan }x-\frac{\beta }{2} {\cosec }x
=\frac{\alpha '_m+\beta }{2}{\cotan }\frac{x}{2}
-\frac{\alpha '_m-\beta }{2}{\tan }\frac{x}{2}\\
\mathcal{V}_m(x)\!=\!\left( {\alpha '_m}^2-\alpha '_m+\frac{\beta ^2}{4} \right) \!{\cosec }^2x
\!-\!(2\alpha '_m\!-\!1)\frac{\beta }{2} {\cotan} x {\cosec }x\!-\!{\alpha '_m}^2\!+\!\lambda _m
\end{array}
\end{equation}
where $\lambda _m=m(m-\alpha -1)$.

\item  {\it Generalized P\"{o}schl-Teller potential}\\
In the case $\sigma (s)\!=\!s^2\!-\!1$, the change of variable
$(0,\infty )\longrightarrow (1,\infty ): x\mapsto s(x)={\cosh }\, x$
leads to 
\begin{equation}
\begin{array}{l}
W_m(x)=\alpha _m {\cotanh }\, x-\frac{\beta }{2} {\cosech }x\\
\mathcal{V}_m(x)\!=\!\left( \alpha _m^2\!+\!\alpha _m\!+\!\frac{\beta ^2}{4} \right) {\cosech }^2x
\!-\!(2\alpha _m\!+\!1)\frac{\beta }{2} {\cotanh }\, x {\cosech }x\!+\!\alpha _m^2\!+\!\lambda _m
\end{array}
\end{equation} 
where $\lambda _m=-m(m+\alpha -1)$.

\item {\it Morse type potential}\\
In the case  $\sigma (s)=s^2$, the change of variable
$\mathbb{R}\longrightarrow (0,\infty ):\, x\mapsto s(x)={\rm e}^x$
leads to 
\begin{equation}
\begin{array}{l}
W_m(x)=-\frac{\beta }{2} {\rm e}^{-x}+\alpha _m\\
\mathcal{V}_m(x)=\frac{\beta ^2}{4}{\rm e}^{-2x}-(2\alpha _m+1)\frac{\beta }{2} {\rm e}^{-x}+\alpha _m^2+\lambda _m
\end{array}
\end{equation}  
where $\lambda _m=-m(m+\alpha -1)$.

\item {\it Scarf hyperbolic type potential}\\
In the case $\sigma (s)\!=\!s^2\!+\!1$, the change of variable
$\mathbb{R}\longrightarrow \mathbb{R}: x\mapsto s(x)={\sinh }\, x$
leads to 
\begin{equation}
\begin{array}{l}
W_m(x)=\alpha _m {\tanh }\, x-\frac{\beta }{2} {\sech }x \\
\mathcal{V}_m(x) =\left(-\alpha _m^2-\alpha _m+\frac{\beta ^2}{4}\right) {\sech }^2x
-(2\alpha _m+1)\frac{\beta }{2} \, {\tanh }\, x {\sech }x +\alpha _m^2+\lambda _m.
\end{array}
\end{equation} 
where $\lambda _m=-m(m+\alpha -1)$.
\end{enumerate}

\section{Other shape invariant operators}

In this section we restrict us \cite{JF} to the particular non-trivial cases when $\alpha $ and $\beta $ 
are such that there exists $k\in \mathbb{R}$ with $\varrho (s)=\sigma ^k(s)$ (see table 2).

\begin{table}[htbp]
\begin{center}
\begin{tabular}{c|llll}
\hline
$\sigma (s)$ & \quad $\tau (s) $  & \quad $\varrho (s)$  & \quad $k$  &  \quad $(a,b) $\\
\hline \hline
$s$ & \quad $\beta $ & \quad $s^{\beta -1}$ & \quad $\beta -1$& $\quad (0,\infty )$\\ 
$1-s^2$  &\quad  $\alpha s$ &\quad  $(1-s^2)^{-\alpha /2-1}$& \quad $-\frac{\alpha }{2}-1$ & \quad $(-1,1)$\\
$s^2-1$  &\quad  $\alpha s$ &\quad  $(s^2-1)^{\alpha /2-1}$ & \quad $\frac{\alpha }{2}-1$ & \quad $(1,\infty )$\\
$s^2$  & $\quad \alpha s$ & \quad $s^{\alpha /2-1}$& \quad $\frac{\alpha }{2}-1$ & \quad $(0,\infty )$\\
$s^2+1$  & \quad $\alpha s$ &\quad  $(s^2+1)^{\alpha /2-1}$ &\quad  $\frac{\alpha }{2}-1$ & \quad $(-\infty ,\infty )$\\
\hline
\end{tabular}
\end{center}
\caption{The cases when $\varrho (s)=\sigma ^k(s)$ }
\end{table}

\noindent From $(\sigma \varrho )'=\tau \varrho $
we get $\tau (s)=(k+1)\sigma '(s)=2(k+1)\kappa (s)\, \kappa '(s)$, and
\begin{equation} \begin{array}{l}
  A_m=\kappa (s)\frac{d}{ds}-m\kappa '(s)\qquad
A_m^+=-\kappa (s)\frac{d}{ds}-(2k+m+1)\kappa '(s)\\[2mm]
{H}_m=-\kappa ^2(s)\frac{d}{ds^2}-2(k+1)\kappa (s)\, \kappa '(s)\frac{d}{ds}-m(m+2k)\kappa (s)\, \kappa ''(s) \\[2mm]
\lambda _m=-m(2k+m+1)\frac{\sigma ''(s)}{2}=-m(2k+m+1)[{\kappa '}^2(s)+\kappa (s)\, \kappa ''(s))].
\end{array} \end{equation}
\begin{theorem}
If $\alpha $ and $\beta $ are such that $\varrho (s)=\sigma ^k(s)$
then for any $\gamma \in \mathbb{R}$ the operators
\begin{equation}
\tilde{A}_m=A_m+\frac{\gamma }{2m+2k+1}\qquad \quad
\tilde{A}_m^+=A_m^++\frac{\gamma }{2m+2k+1} 
\end{equation}
satisfy for $m<\Lambda -1$ with $2m+2k+1\not=0$ the relations
\begin{equation} \label{shapeinvar2} 
\begin{array}{ll}
\tilde{A}_m^+\tilde{A}_m=\tilde{H}_m-\tilde{\lambda }_m\qquad & 
\tilde{A}_m\tilde{H}_m=\tilde{H}_{m+1}\tilde{A}_m\\[2mm]
\tilde{A}_m\tilde{A}_m^+=\tilde{H}_{m+1}-\tilde{\lambda }_m \qquad &
\tilde{H}_m\tilde{A}_m^+=\tilde{A}_m^+\tilde{H}_{m+1}
\end{array}
\end{equation}
where 
\begin{equation}
\tilde{H}_m={H}_m-\gamma \frac{d\kappa }{ds}\qquad \qquad 
\tilde{\lambda }_m=\lambda _m-\frac{\gamma ^2}{(2m+2k+1)^2}.
\end{equation}
\end{theorem}
\begin{proof*} 
Since $A_m^+A_m={H}_m-\lambda _m$ and $A_mA_m^+={H}_{m+1}-\lambda _m$ we obtain
\[ \begin{array}{l}
(A_m^++\varepsilon )(A_m+\varepsilon )={H}_m-\lambda _m-\varepsilon (2m+2k+1)\kappa '(s)+\varepsilon ^2 \\
(A_m+\varepsilon )(A_m^++\varepsilon )={H}_{m+1}-\lambda _m-\varepsilon (2m+2k+1)\kappa '(s)+\varepsilon ^2 
\end{array} \]
for any constant $\varepsilon $. If we choose $\varepsilon =1/(2m+2k+1)$ then we get (\ref{shapeinvar2})
\[ \begin{array}{l}
\tilde{H}_m\tilde{A}_m^+=(\tilde{A}_m^+\tilde{A}_m+\tilde{\lambda }_m)\tilde{A}_m^+=
\tilde{A}_m^+(\tilde{A}_m\tilde{A}_m^++\tilde{\lambda }_m)=\tilde{A}_m^+\tilde{H}_{m+1}\\
\tilde{A}_m\tilde{H}_m=\tilde{A}_m(\tilde{A}_m^+\tilde{A}_m+\tilde{\lambda }_m)=
(\tilde{A}_m\tilde{A}_m^++\tilde{\lambda }_m)\tilde{A}_m=\tilde{H}_{m+1}\tilde{A}_m.
\end{array} \]
\end{proof*}
\begin{theorem}
If \ $0\!\leq \!m \!\leq \!l<\!\Lambda $ and if \ $\tilde{\Phi }_{l,l}$ \ satisfies the relation 
$\tilde{A}_l\tilde{\Phi }_{l,l}=0$ then
\begin{equation} \label{tildephi}
\tilde{\Phi }_{l,m}=
\frac{\tilde{A}_m^+ }{\tilde{\lambda }_l-\tilde{\lambda }_m}\
\frac{\tilde{A}_{m+1}^+ }{\tilde{\lambda }_l-\tilde{\lambda }_{m+1}}\ ...\
\frac{\tilde{A}_{l-2}^+ }{\tilde{\lambda }_l-\tilde{\lambda }_{l-2}}\
\frac{\tilde{A}_{l-1}^+ }{\tilde{\lambda }_l-\tilde{\lambda }_{l-1}}\, \tilde{\Phi }_{l,l}
\end{equation}
is an eigenfunction of $\tilde{H}_m$
\begin{equation}\label{schrod2}
\tilde{H}_m\tilde{\Phi }_{l,m}=\tilde{\lambda }_l\tilde{\Phi }_{l,m}
\end{equation}
and 
\begin{equation}\begin{array}{l}
\tilde{A}_m\tilde{\Phi }_{l,m}=\left\{ \begin{array}{lll}
0 & {\rm if} & m=l\\
\tilde{\Phi }_{l,m+1} & {\rm if} & m<l
\end{array} \right. \\
\tilde{A}_m^+\tilde{\Phi }_{l,m+1}=(\tilde{\lambda }_l-\tilde{\lambda }_m)\tilde{\Phi }_{l,m}.
\end{array}
\end{equation}
\end{theorem}
\begin{proof*} 
The definition (\ref{tildephi}) of $\tilde{\Phi }_{l,m}$ can be re-written as
\begin{equation} \label{rectildephi}
\tilde{\Phi }_{l,m}=
\frac{\tilde{A}_m^+ }{\tilde{\lambda }_l-\tilde{\lambda }_m}\, \tilde{\Phi }_{l,m+1}
\end{equation}
and $\tilde{H}_l\tilde{\Phi }_{l,l}
\!=\!(\tilde{A}_l^+\tilde{A}_l+\tilde{\lambda }_l)\tilde{\Phi }_{l,l}\!=\!\tilde{\lambda }_l\tilde{\Phi }_{l,l}$.
The relation $\tilde{H}_m\tilde{\Phi }_{l,m}\!=\!\tilde{\lambda }_l\tilde{\Phi }_{l,m}$ follows by recurrence
\[   
\tilde{H}_{m+1}\tilde{\Phi }_{l,m+1}\!=\!\tilde{\lambda }_l\tilde{\Phi }_{l,m+1}\quad \Longrightarrow \quad 
\tilde{H}_m\tilde{\Phi }_{l,m}
\!=\!\frac{\tilde{H}_m\tilde{A}_m^+ }{\tilde{\lambda }_l-\tilde{\lambda }_m}\, \tilde{\Phi }_{l,m+1}
\!=\!\frac{\tilde{A}_m^+ \tilde{H}_{m+1} }{\tilde{\lambda }_l-\tilde{\lambda }_m}\, \tilde{\Phi }_{l,m+1}
\!=\!\tilde{\lambda }_l\tilde{\Phi }_{l,m}.
\]
From the relation (\ref{rectildephi}) we get
\[
\tilde{A}_m\tilde{\Phi }_{l,m}
=\frac{\tilde{A}_m\tilde{A}_m^+ }{\tilde{\lambda }_l-\tilde{\lambda }_m}\, \tilde{\Phi }_{l,m+1}
=\frac{\tilde{H}_{m+1}-\tilde{\lambda }_m}{\tilde{\lambda }_l-\tilde{\lambda }_m}\, \tilde{\Phi }_{l,m+1}
=\tilde{\Phi }_{l,m+1}.\]
\end{proof*}

If in equation (\ref{schrod2}) we pass to a new variable $x=x(s)$ such that
$dx/ds=\pm 1/\kappa (s)$ and to the new functions 
\begin{equation}
\tilde{\Psi }_{l,m}(x)=\sqrt{\kappa (s(x))\, \varrho (s(x))}\, \tilde{\Phi }_{l,m}(s(x)).
\end{equation}
then we get the Schr\" odinger type equation 
\begin{equation}
\tilde{\mathcal{H}}_m\tilde{\Psi }_{l,m}=\tilde{\lambda }_l\tilde{\Psi }_{l,m}
\end{equation}
where
\begin{equation}
\tilde{\mathcal{H}}_m=-\frac{d^2}{dx^2}+\tilde{\mathcal{V}}_m(x)\qquad {\rm and}\qquad 
\tilde{\mathcal{V}}_m(x)=\mathcal{V}_m(x)-\gamma \frac{d\kappa }{ds}(s(x)).
\end{equation}
 
The operators
\begin{equation}\label{tildeA+}
 \begin{array}{l}
\tilde{\mathcal A}_m=[\kappa (s)\varrho (s)]^{1/2}\tilde{A}_m[\kappa (s)\varrho (s)]^{-1/2}|_{s=s(x)}\\[2mm]
\tilde{\mathcal A}_m^+ =[\kappa (s)\varrho (s)]^{1/2}\tilde{A}_m^+ [\kappa (s)\varrho (s)]^{-1/2}|_{s=s(x)}
\end{array}
\end{equation}
corresponding to $\tilde{A}_m$ and $\tilde{A}_m^+ $ satisfy the relations
\begin{equation} \begin{array}{l}
  \tilde{\mathcal A}_m\tilde{\Psi }_{l,m}(x)=\left\{ \begin{array}{lll}
0 & {\rm if} & m=l\\
\tilde{\Psi }_{l,m+1} & {\rm if} & m<l
\end{array} \right. \\[5mm]
\tilde{\mathcal A}_m^+ \tilde{\Psi }_{l,m+1}(x)=(\tilde{\lambda }_l-\tilde{\lambda }_m)\tilde{\Psi }_{l,m}(x)
\end{array}
\end{equation}
and
\begin{equation}
\begin{array}{ll}
\tilde{\mathcal{H}}_m-\tilde{\lambda }_m=\tilde{\mathcal{A}}_m^+\tilde{\mathcal{A}}_m & 
\tilde{\mathcal{A}}_m\tilde{\mathcal{H}}_m=\tilde{\mathcal{H}}_{m+1}\tilde{\mathcal{A}}_m \\[2mm]
\tilde{\mathcal{H}}_{m+1}-\tilde{\lambda }_m =\tilde{\mathcal{A}}_m\tilde{\mathcal{A}}_m^+\qquad  & 
\tilde{\mathcal{H}}_m\tilde{\mathcal{A}}_m^+=\tilde{\mathcal{A}}_m^+\tilde{\mathcal{H}}_{m+1}. 
\end{array}
\end{equation} 

If the change of variable $s=s(x)$ is such that $ds/dx=\pm \kappa (s(x))$ 
then
\begin{equation}
\tilde{\mathcal A}_m=\pm \frac{d}{dx}+\tilde{W}_m(x)
\qquad \qquad \tilde{\mathcal A}_m^+ =\mp \frac{d}{dx}+\tilde{W}_m(x)
\end{equation}
\begin{equation}
\tilde{\mathcal{V}}_m(x)
=\tilde{W}_m^2(x)\mp \dot {\tilde{W}}_m(x)+\tilde{\lambda }_m 
=\frac{\ddot {\tilde{\Psi }}_{m,m}(x)}{\tilde{\Psi }_{m,m}(x)}+\tilde{\lambda }_m 
\end{equation}
where $\tilde{W}_m(x)$ is the superpotential \cite{JF}
\begin{equation}\label{Wm}
\tilde{W}_m(x)=-\frac{\tau (s(x))}{2\kappa (s(x))}
- \left(m-\frac{1}{2}\right )\frac{d\kappa }{ds}(s(x)) +\frac{\gamma }{2m+2k+1}
=\mp \frac{\dot {\tilde{\Psi }}_{m,m}(x)}{\tilde{\Psi }_{m,m}(x)}.
\end{equation}

\noindent {\bf Particular cases} \cite{CKS,D,JF}.
Let $\alpha _m=-(2m+\alpha -1)/2$, \ $\alpha '_m=(2m-\alpha -1)/2$.
\begin{enumerate}

\item {\it Coulomb type potential}\\
In the case $\sigma (s)=s$, the change of variable
$(0,\infty )\longrightarrow (0,\infty ):\, x\mapsto s(x)=x^2/4$
leads to 
\begin{equation}
\begin{array}{l}
\tilde{W}_m(x)=-\left(\beta +m-\frac{1}{2}\right) \frac{1}{x}+\frac{\gamma }{2m+2\beta -1}\\
\tilde{\mathcal{V}}_m(x)=\left(\beta +m-\frac{1}{2}\right)\left(\beta +m-\frac{3}{2}\right)\frac{1}{x^2}
-\gamma \frac{1}{x}\\
\tilde{\lambda }_m=-\frac{{\gamma }^2}{(2m+2\beta -1)^2}.\\
\end{array}
\end{equation}

\item {\it Trigonometric Rosen-Morse type potential}\\  
In the case $\sigma (s)=1-s^2$, the change of variable
$(0,\pi )\longrightarrow (-1,1):\, x\mapsto s(x)=\cos x$
leads to 
\begin{equation}
\begin{array}{l}
\tilde{W}_m(x)=\alpha '_m{\cotan }\, x+\frac{\gamma }{2m-\alpha -1}\\
\tilde{\mathcal{V}}_m(x)=\left( {\alpha '_m}^2-\alpha '_m\right){\cosec }^2x+
\gamma \, {\cotan }x-{\alpha '_m}^2+m(m-\alpha -1)\\
\tilde{\lambda }_m=m(m-\alpha -1)-\frac{{\gamma }^2}{(2m-\alpha -1)^2}
\end{array} 
\end{equation}

\item  {\it Eckart type potential}\\
In the case $\sigma (s)\!=\!s^2\!-\!1$, the change of variable
$(0,\infty )\longrightarrow (1,\infty ):\, x\mapsto s(x)={\cosh }\, x$
leads to 
\begin{equation}
\begin{array}{l}
\tilde{W}_m(x)=\alpha _m {\cotanh }\, x+\frac{\gamma }{2m+\alpha -1}\\
\tilde{\mathcal{V}}_m(x)=\left( \alpha _m^2+\alpha _m\right) {\cosech }^2x-
\gamma {\cotanh }x+\alpha _m^2-m(m-\alpha -1)\\
\tilde{\lambda }_m=-m(m-\alpha -1)-\frac{{\gamma }^2}{(2m+\alpha -1)^2}.
\end{array}
\end{equation} 

\item {\it Hyperbolic Rosen-Morse type potential}\\
In the case $\sigma (s)\!=\!s^2\!+\!1$, the change of variable
$\mathbb{R}\!\longrightarrow \!\mathbb{R}:\, x\mapsto s(x)\!=\!{\sinh }\, x$
leads to 
\begin{equation}
\begin{array}{l}
\tilde{W}_m(x)=\alpha _m {\tanh }\, x+\frac{\gamma }{2m+\alpha -1}\\
\tilde{\mathcal{V}}_m(x) =-\left(\alpha _m^2+\alpha _m\right){\sech }^2x 
-\gamma \, {\tanh }\, x +\alpha _m^2-m(m-\alpha -1)\\
\tilde{\lambda }_m=-m(m-\alpha -1)-\frac{{\gamma }^2}{(2m+\alpha -1)^2}.
\end{array}
\end{equation} 
\end{enumerate}

\section{Concluding remarks}

Most of the known exactly solvable Schr\" odinger equations are directly related to some 
shape invariant operators, and most of the formulae occurring in the study of these quantum 
systems follow from a small number of mathematical results concerning the hypergeometric type operators. 
It is simpler to study these shape invariant operators then the corresponding operators 
occurring in various applications to quantum mechanics. Our systematic study recovers known
results in a natural unified way, and allows one to extend certain results known in particular cases.

\section*{Acknowledgment}

This  research was supported by the grant CEx06-......

\end{document}